\documentclass[lettersize,journal]{IEEEtran}
\usepackage{amsmath,graphicx,mathtools,physics}
\usepackage{lineno}
\usepackage{bm}

\usepackage{amsmath,amsfonts}
\usepackage{algorithmic}
\usepackage{algorithm}
\usepackage{array}
\usepackage[caption=false,font=normalsize,labelfont=sf,textfont=sf]{subfig}
\usepackage{textcomp}
\usepackage{stfloats}
\usepackage{url}
\usepackage{verbatim}
\usepackage{graphicx}
\usepackage{cite}
\begin{document}

\title{Three-Dimensional FDTD Simulation of Superconducting Terahertz Emitters Using an Effective Medium Approximation} 
\author{Ryota Kobayashi\thanks{Department of Electronic Science and Engineering, Graduate School of Engineering, Kyoto University, Kyoto 615-8510, Japan}, Itsuhiro Kakeya\thanks{Department of Electronic Science and Engineering, Graduate School of Engineering, Kyoto University, Kyoto 615-8510, Japan}}




\maketitle

\begin{abstract}
Compact and tunable terahertz (THz) sources with controllable polarization are sought for high-speed links, spectroscopy, and imaging. In superconducting THz emitters based on a cuprate  Bi$_2$Sr$_2$CaCu$_2$O$_{8+\delta}$, which have the potential to satisfy such properties, modeling often relies on heavy nonlinear solvers or empirical isotropic dielectrics, obscuring anisotropy, plasma response, and bias dependence. We derive an anisotropic, dispersive effective permittivity from the Lawrence-Doniach model for a layered superconductor and implement it as a user-defined material in a commercial Finite-difference Time-domain (FDTD) solver. Simulations of two device geometries reproduce linearly polarized emission from a rectangular mesa and low-axial-ratio (near-circular) emission from a truncated square mesa. This CAD-ready workflow links superconducting theory to device design, enabling rapid searches and parameter optimization for next-generation THz sources.
\end{abstract}


\section{Introduction}
The Josephson Plasma Emitter (JPE) is a terahertz source that uses the bismuth-based compound $\mathrm{Bi_2Sr_2CaCu_2O_{8+\delta}}$ (Bi2212), a typical cuprate high-temperature superconductor \cite{Ozyuzer2007-nq,Kadowaki2008-rs,Hu2010-tv,Welp2013-eb,Kakeya2016-ns,Kleiner2019-hy,Delfanazari2025-ys}. This source utilizes the AC Josephson effect, in which an alternating superconducting current flows at a frequency proportional to the voltage applied across the multiple layers of Josephson junctions (intrinsic Josephson junctions) present in the Bi2212 crystal structure, and the resonance from a micro-cavity structure (mesa structure) formed on the crystal surface. First demonstrated by Ozyuzer et al. in 2007 \cite{Ozyuzer2007-nq}, the JPE has attracted attention for its novel operating principle and its applications as a terahertz imaging device \cite{Tsujimoto2012-le}. A notable feature of the JPE is the wide frequency tunability of its emitted radiation based on the AC Josephson effect; for instance, radiation from 0.5 to 2.4 THz has been observed from a single device \cite{Kashiwagi2015-og}.

This frequency tunability also holds for high-frequency modulated voltages. Recently, Miyamoto et al. confirmed through spectral observation of the emitted terahertz waves that the JPE maintains the voltage-frequency relationship described by the AC Josephson effect even for a 3 GHz modulation voltage \cite{Miyamoto2024-el}. Combined with monolithic polarization control through mesa geometry \cite{Elarabi2017-uk,Elarabi2024-np}, these multifunctional capabilities make JPEs attractive candidates for compact terahertz source systems serving communications, spectroscopy, and imaging.

Translating these capabilities into practical devices, however, requires systematic design of the coupling between the mesa and external structures---antennas, waveguides, and optical elements---that collect and shape the emitted field. Such co-design demands device models that faithfully represent the electromagnetic behavior of the mesa while remaining compatible with general-purpose electromagnetic simulators. Yet this integration has remained limited.

Numerical analyses of JPEs have examined thermal distributions \cite{Yurgens2011-fr,Gross2012-vn} and electromagnetic-field distributions \cite{Koyama2009-an,Hu2009-yo,Machida2010-jw,Gross2015-zi,Rudau2016-ov}, typically using researcher-built Finite-difference Time-domain (FDTD) codes to capture junction nonlinearity and superconducting electrodynamics. These efforts revealed mechanisms such as localized high-temperature "hot spots" \cite{Yurgens2011-fr,Gross2012-vn} and $\pi$-kink resonant modes \cite{Hu2009-yo}. Yet, homegrown codes often lack speed, modeling flexibility, and easy sharing, so they have aided principal verification more than device design. For design-oriented work, researchers usually prioritize accurate electromagnetic modes and resonances over rigorous superconducting dynamics—for example, approximating Bi2212 as an isotropic dielectric with refractive index $\sim 4$ and using commercial FDTD tools \cite{Elarabi2016-lr,Saito2022-fw}. This bold simplification enables high-performance calculations and reproduces key characteristics (resonance, polarization). Still, its theoretical basis is unclear because it discards not only Josephson nonlinearity but also linear features, such as plasma resonance and anisotropy of layered crystals, making it essentially an empirical, experiment-matching approach.

In this paper, we derive an anisotropic dielectric model for a general FDTD simulator from the rigorous mathematical model of a JPE using linear and effective medium approximations. Next, we report that we implemented the derived anisotropic dielectric model in a commercial FDTD simulator and reproduced the previously observed resonance and polarization characteristics of multiple devices. This method improves on the empirical isotropic dielectric model that experimental researchers have used in the past, and its theoretical basis is ensured by its derivation from the rigorous mathematical model of a stacked Josephson junction system. In addition, although this model cannot reproduce the nonlinear characteristics of the Josephson effect, it can reproduce linear characteristics, such as plasma resonance. Since it operates on a commercial FDTD simulator, flexible modeling and high-speed numerical calculations are possible, backed by the simulator's performance.

\section{Formulations}
The anisotropic dielectric model of cuprate high-temperature superconductors has been used to perform band calculations of negative refraction \cite{Rakhmanov2010-xp}. In this paper, we derive this model in a rigorous form from the Lawrence-Doniach (LD) model \cite{Lawrence1971-tv}, a semi-classical model describing a stacked Josephson junction system, and extend the discussion to the case of a finite applied voltage. The details of the derivation are described in the supplementary material.

Here, we give a brief outline of the derivation of the anisotropic dielectric model. The starting point of the discussion is to regard the layered structure of Bi2212 as a kind of hyperbolic metamaterial \cite{Smith2003-ji,Poddubny2013-wg}. This suggests that if the in-plane and out-of-plane complex permittivities in each layer are known, the effective permittivity tensor $\tilde{\bar{\varepsilon}}_{\rm eff}$ for the entire bulk can be written as
\begin{gather}
    \tilde{\bar{\varepsilon}}_{\rm eff}=
    \begin{pmatrix}
        \tilde\varepsilon_{\rm eff}^\parallel & 0 & 0 \\
        0 & \tilde\varepsilon_{\rm eff}^\parallel & 0 \\
        0 & 0 & \tilde\varepsilon_{\rm eff}^\perp
    \end{pmatrix},\\
    \tilde\varepsilon_{\rm eff}^\parallel=\frac{D_{\rm S}\tilde\varepsilon_{\rm S}^\parallel+D_{\rm I}\tilde\varepsilon_{\rm I}^\parallel}{D},\;\;\;\;
    \tilde\varepsilon_{\rm eff}^\perp=\frac{\tilde\varepsilon_{\rm S}^\perp\tilde\varepsilon_{\rm I}^\perp D}{D_{\rm S}\tilde\varepsilon_{\rm I}^\perp+D_{\rm I}\tilde\varepsilon_{\rm S}^\perp},
\label{meq:effective_permittivity_tensor}
\end{gather}
where $\tilde\varepsilon_{\rm S}^\parallel$, $\tilde\varepsilon_{\rm I}^\parallel$, $\tilde\varepsilon_{\rm S}^\perp$, and $\tilde\varepsilon_{\rm I}^\perp$ are the in-plane and out-of-plane complex permittivities in the superconducting and insulating layers, respectively. The LD model directly determines the superconducting-layer in-plane response and the insulating-layer out-of-plane response. Using the Fourier convention $\exp(-i\omega t)$, the dissipative contribution enters with a positive imaginary part. In the unbiased state, the $q_z=0$ out-of-plane response and, after taking $\tilde\varepsilon_{\rm I}^\parallel\approx\varepsilon_{\rm I}$, the resulting in-plane effective permittivity are
\begin{align}
    \varepsilon_0\tilde\varepsilon_{\rm eff}^\parallel
        &=\varepsilon_0\varepsilon_\infty^\parallel
          -\varepsilon_0\varepsilon_{\rm I}\frac{\Gamma^2\omega_{\rm p}^2}{\omega^2}
          +i\frac{\sigma_\parallel}{\omega},\\
    \varepsilon_0\tilde\varepsilon_{\rm I}^\perp
        &=\varepsilon_0\varepsilon_{\rm I}
          -\varepsilon_0\varepsilon_{\rm I}\frac{\omega_{\rm p}^2}{\omega^2}
          +i\frac{\sigma_{\rm I}}{\omega}.
\label{meq:oop_complex_permittivity_in_I_layer}
\end{align}
Here, $\varepsilon_0$ is the permittivity of vacuum, $\varepsilon_{\rm S}$ and $\varepsilon_{\rm I}$ are the relative permittivities of the superconducting and insulating layers, respectively, $\sigma_{\rm S}$ and $\sigma_{\rm I}$ are the conductivities of the superconducting and insulating layers, respectively, $\omega_{\rm p}$ is the Josephson plasma frequency, $\Gamma$ is the superconducting anisotropy parameter, $\varepsilon_\infty^\parallel \equiv (D_{\rm S}\varepsilon_{\rm S}+D_{\rm I}\varepsilon_{\rm I})/D$, and $\sigma_\parallel \equiv (D_{\rm S}/D)\sigma_{\rm S}$. The remaining two local permittivities, $\tilde\varepsilon_{\rm S}^\perp$ and $\tilde\varepsilon_{\rm I}^\parallel$, are not determined by the LD model. For the out-of-plane closure used in the simulations below, we assume that the superconducting-layer term in the harmonic average is negligible, namely $|(D_{\rm S}/D)/\tilde\varepsilon_{\rm S}^\perp|\ll|(D_{\rm I}/D)/\tilde\varepsilon_{\rm I}^\perp|$, equivalently $|D_{\rm I}\tilde\varepsilon_{\rm S}^\perp|\gg|D_{\rm S}\tilde\varepsilon_{\rm I}^\perp|$. Under this condition, Eq. \eqref{meq:effective_permittivity_tensor} reduces to $\tilde\varepsilon_{\rm eff}^\perp\approx(D/D_{\rm I})\tilde\varepsilon_{\rm I}^\perp$ in the unbiased and zero-voltage states; for example, this condition is satisfied when $|\tilde\varepsilon_{\rm S}^\perp|\gg|\tilde\varepsilon_{\rm I}^\perp|$. The more general $q_z$-dependent response and the approximation details are given in the supplementary material.

The derived complex effective permittivity tensor clearly captures the physical picture of the high-frequency electric-field response of Bi2212. The term proportional to $\omega^{-2}$ in the in-plane effective permittivity $\tilde\varepsilon_{\rm eff}^\parallel$ represents the inertial response of Cooper pairs in the superconducting layers, i.e., their inductive behavior. This arises from the phase lag in the motion of Cooper pairs relative to the external electric field, leading the material to behave as a plasma with negative permittivity.

The out-of-plane effective permittivity $\tilde\varepsilon_{\rm eff}^\perp$ changes its properties significantly depending on the bias state of the system. In the zero-voltage state, Cooper pair tunneling in the Josephson junctions is dominant, and a plasma-like response due to Josephson plasma resonance is observed. However, when a finite DC voltage is applied, the Josephson junction enters the AC Josephson regime and oscillates at a high frequency. In that regime, the linear response is formally described by a Floquet matrix because frequency mixing couples $\omega$ to $\omega\pm\omega_{\rm B}$. Under the diagonal approximation used here, in which the sideband electric fields are neglected, only the capacitive background and normal conductive loss of the insulating layers remain. In that approximation, $\tilde\varepsilon_{\rm I}^\perp$ is different from Eq. \eqref{meq:oop_complex_permittivity_in_I_layer} and is given by
\begin{align}
    \varepsilon_0\tilde\varepsilon_{\rm I}^{\perp,\mathrm{diag}}
        =\varepsilon_0\varepsilon_{\rm I}+i\frac{\sigma_{\rm I}}{\omega}.
\end{align}
For details on this derivation and its range of validity, please refer to the supplementary material.

In both the in-plane and out-of-plane directions, the imaginary part of the permittivity represents the conductive loss due to the scattering of normal electrons. This model thus expresses the anisotropic and dynamic dielectric properties of the JPE with a single complex permittivity tensor.

\section{Simulation Method}
The complex effective permittivity tensor $\tilde{\bar{\varepsilon}}_{\rm eff}$ of Bi2212, obtained as described above, was entered as a material parameter in the FDTD simulation software, CST Studio Suite, Dassault Syst\'{e}mes. CST Studio Suite was run on the laboratory's workstation, a Dell Precision 3450 (Xeon W1290, 64 GB RAM). We modeled the following two devices, which our group has previously measured, in CST Studio Suite and compared the simulation results with the experimental results.

\subsection{Rectangular mesa device}
In Device R1 (Figure \ref{fig:devices_and_models} (a)), intensive radiations from mesas B, C, and E were reported \cite{Kobayashi2022-px}. In the present simulation, the terahertz radiation from mesa B was reproduced. The modeled mesa has a short side of 70 $\mu$m, a long side of 350 $\mu$m, and a thickness of 1.2 $\mu$m. To reflect the characteristics of mesa formation by Ar ion milling, the sidewall inclination is set to $26.5^\circ$ (Figure \ref{fig:devices_and_models} (b)).

\subsection{Truncated Square device}
\begin{figure}
    \centering
    \includegraphics[width=\linewidth]{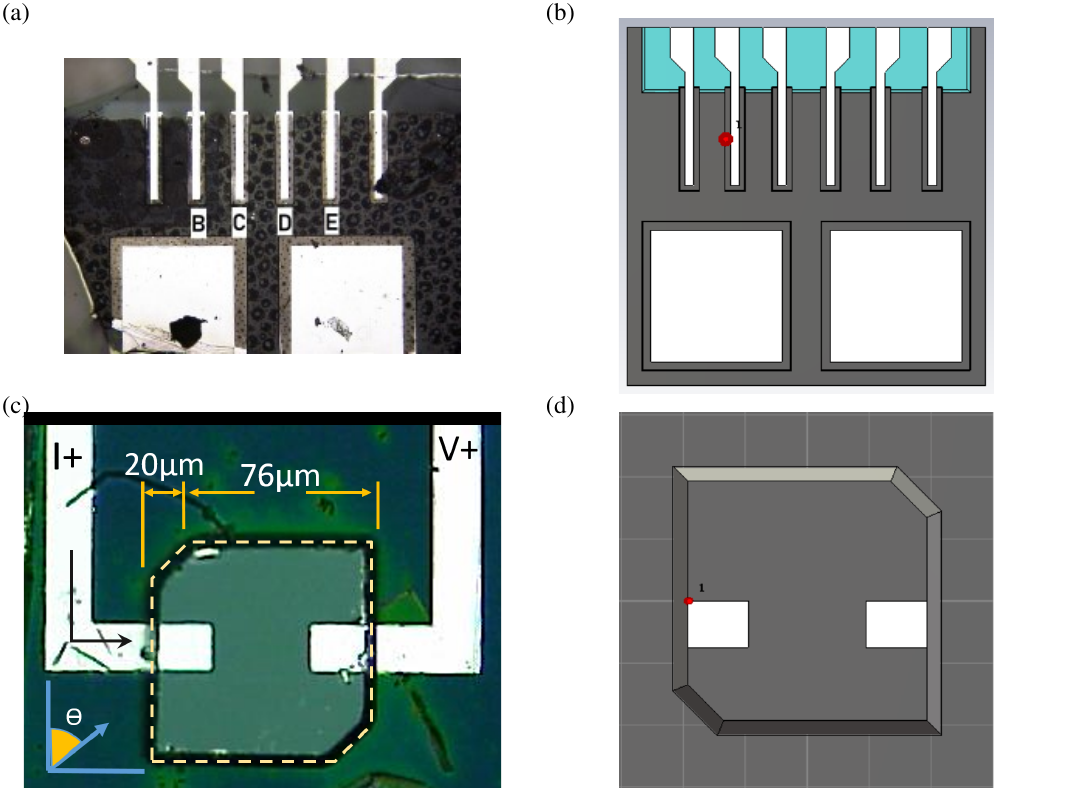}
    \caption[Optical micrographs and CST models of the two mesa devices]{
    (a) Optical micrograph of the rectangular-mesa device (Device R1). Individual mesas are labeled, and strong emissions were previously reported from mesas B, C, and E; in this work, we model mesa B.
    (b) CST Studio Suite model of Device R1 used for 3D FDTD analysis. The mesa geometry is set to $70\,\mu\mathrm{m}\times 350\,\mu\mathrm{m}$ with a thickness of $1.2\,\mu\mathrm{m}$ and sidewalls inclined by $26.5^{\circ}$ to reflect Ar ion milling; the red dot marks the current-excitation port at the mesa edge.
    (c) Optical micrograph of the truncated-square device (Device TS) with bias/voltage pads (\mbox{I$+$}, \mbox{V$+$}) indicated. The designed mesa is an $82\,\mu\mathrm{m}$ square with a pair of opposite corners truncated by $13\,\mu\mathrm{m}$ (the device in the photo is a representative device of the same shape but with different dimensions).
    (d) CST model of Device TS with the same $26.5^{\circ}$ sidewall inclination and a thickness of $2.4\,\mu\mathrm{m}$; the red dot indicates the edge excitation used in the simulations.}
    \label{fig:devices_and_models}
\end{figure}

In Device TS (Figure \ref{fig:devices_and_models} (c)), terahertz electromagnetic wave radiations with low axial ratio (AR) were measured \cite{Elarabi2017-uk}. The obtained AR strongly depends on the radiation frequency. The modeled mesa has a square shape with a side of 82 $\mu$m, a pair of opposite corners cut by 13 $\mu$m, and a thickness of 2.4 $\mu$m. As with device R1, the sidewall inclination is set to $26.5^\circ$ to reflect the characteristics of mesa formation by Ar ion milling (Figure \ref{fig:devices_and_models} (d)).

\subsection{Excitation and Observation}
\begin{figure}
    \centering
    \includegraphics[width=\linewidth]{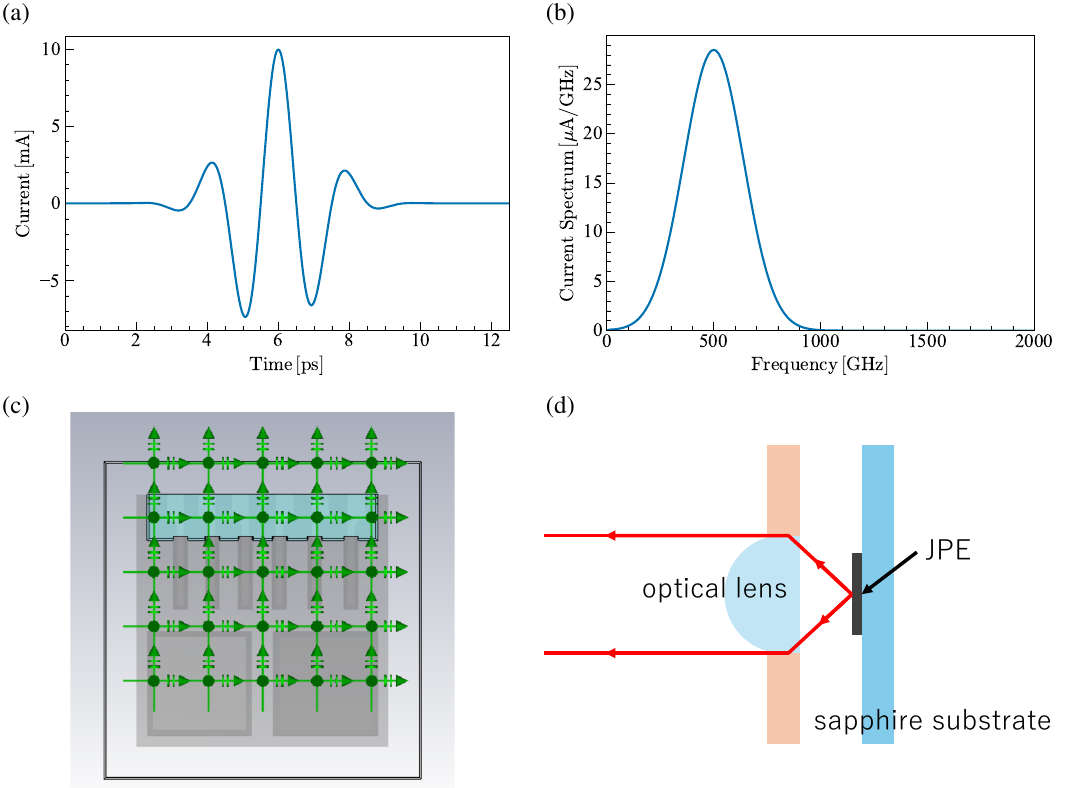}
    \caption[Excitation waveform/spectrum and field-probe arrangement]{
    (a) Time-domain waveform of the injected Gaussian pulse current applied at the mesa edge in the simulations.
    (b) Frequency spectrum of the excitation; within the $2\sigma$ interval it covers approximately $500\pm 300$~GHz, enabling broadband modal excitation around the target band.
    (c) Arrangement of far-field electric- and magnetic-field probes: a square grid with $0.25\,\mathrm{mm}$ pitch placed $0.5\,\mathrm{mm}$ above the mesa. The time-averaged Poynting flux is integrated over the observation plane as defined in Eq.~\eqref{meq:radiated_power}, emulating lens collection.
    (d) Schematic of the experimental collection method: terahertz radiation from the JPE on a sapphire substrate is gathered by an optical lens and detected by a bolometer; the simulation post-processing mirrors this collection to enable direct comparison.}
    \label{fig:excitation_and_measurement}
\end{figure}

Since this model is purely linear, it is necessary to provide external excitation to observe the resonance characteristics. In this study, a Gaussian pulse current with a time waveform as shown in Figure \ref{fig:excitation_and_measurement} (a) was exerted at the edge of the mesa. The frequency spectrum of this signal is as shown in Figure \ref{fig:excitation_and_measurement} (b), and it includes a range of $\pm$300 GHz around 500 GHz in the $2\sigma$ interval. This technique can excite various modes depending on the excitation point, but here we focus on the dominant resonant mode exhibiting the desired polarization characteristics.

In this study, far-field electric- and magnetic-field probes arranged in a grid at 0.25 mm intervals were placed 0.5 mm directly above the mesa (Figure \ref{fig:excitation_and_measurement} (c)). From the Fourier-transformed probe fields $\tilde{\bm{E}}_n(f)$ and $\tilde{\bm{H}}_n(f)$ at the $n$-th probe ($n=1,\dots,N$) and the port-current spectrum $\tilde{I}_{\mathrm{port}}(f)$, we defined the normalized fields
\begin{align}
    \hat{\bm{E}}_n(f)=\tilde{\bm{E}}_n(f)\frac{I_0}{\tilde{I}_{\mathrm{port}}(f)},\quad
    \hat{\bm{H}}_n(f)=\tilde{\bm{H}}_n(f)\frac{I_0}{\tilde{I}_{\mathrm{port}}(f)},
\label{meq:normalized_fields}
\end{align}
where $I_0$ is the reference current amplitude. This normalization removes the excitation-spectrum dependence, so that the resulting spectra represent the intrinsic resonance response of the mesa. We then computed the time-averaged Poynting flux normal to the observation plane:
\begin{align}
    S_{z,n}(f)
    = \frac{1}{2}\Re\!\left\{
        \hat{E}_{x,n}(f)\hat{H}_{y,n}^*(f)
        -\hat{E}_{y,n}(f)\hat{H}_{x,n}^*(f)
    \right\},
\label{meq:poynting_flux}
\end{align}
where ${}^*$ denotes the complex conjugate. The radiated power collected over the observation plane was obtained by numerical integration:
\begin{align}
    P_{\mathrm{sim}}(f)
    = \sum_{n=1}^{N} S_{z,n}(f)\,\Delta A,
\label{meq:radiated_power}
\end{align}
where $\Delta A=\Delta x\,\Delta y$ is the area element corresponding to the probe spacing. This emulates the situation in a real experiment, where terahertz light from the JPE is collected by an optical lens placed directly above the mesa and detected by a bolometer (Figure \ref{fig:excitation_and_measurement} (d)).

The polarization of the radiated wave was characterized using the Stokes parameters \cite{Elarabi2024-np}. For each probe, the Stokes parameters are
\begin{align}
    S_{0,n}(f) &= |\hat{E}_{x,n}(f)|^2 + |\hat{E}_{y,n}(f)|^2,\label{meq:s0_at_each_probe}\\
    S_{1,n}(f) &= |\hat{E}_{x,n}(f)|^2 - |\hat{E}_{y,n}(f)|^2,\label{meq:s1_at_each_probe}\\
    S_{2,n}(f) &= 2\Re\!\left\{\hat{E}_{x,n}(f)\hat{E}_{y,n}^*(f)\right\},\label{meq:s2_at_each_probe}\\
    S_{3,n}(f) &= -2\Im\!\left\{\hat{E}_{x,n}(f)\hat{E}_{y,n}^*(f)\right\}.\label{meq:s3_at_each_probe}
\end{align}
We defined the normalized Stokes parameters $s_{j,n}(f)=S_{j,n}(f)/S_{0,n}(f)$ and computed the plane-integrated Stokes spectral density weighted by the local Poynting flux:
\begin{align}
    \bar{S}_j(f) = \sum_{n=1}^{N} s_{j,n}(f)\, S_{z,n}(f)\,\Delta A
    \qquad (j=0,1,2,3).
\label{meq:def:averaged_stokes_parameter}
\end{align}
By construction, $\bar{S}_0(f)=P_{\mathrm{sim}}(f)$. The polarized component intensity is
\begin{align}
    \bar S_{0\rm p}(f)=\sqrt{\bar S_1^2(f)+\bar S_2^2(f)+\bar S_3^2(f)}.
\end{align}
The Jones vector $\bm E_{\rm J}(f)$ is reconstructed as
\begin{align}
    \bm E_{\rm J}(f)&=
    \begin{pmatrix}
        \bar E_{0x}(f)\\
        \bar E_{0y}(f)\exp{i\bar\delta(f)}
    \end{pmatrix},\label{meq:Jones_vector}\\
    \bar E_{0x}(f)&=\sqrt{\frac{\bar S_{0\rm p}(f)+\bar S_1(f)}{2}},\label{meq:e0x}\\
    \bar E_{0y}(f)&=\sqrt{\frac{\bar S_{0\rm p}(f)-\bar S_1(f)}{2}},\label{meq:e0y}\\
    \bar\delta(f)&=\arctan{\frac{\bar S_3(f)}{\bar S_2(f)}}.\label{meq:delta}
\end{align}

\section{Results}
\subsection{Rectangular device}
\begin{figure}[htbp]
    \centering
    \includegraphics[width=\linewidth]{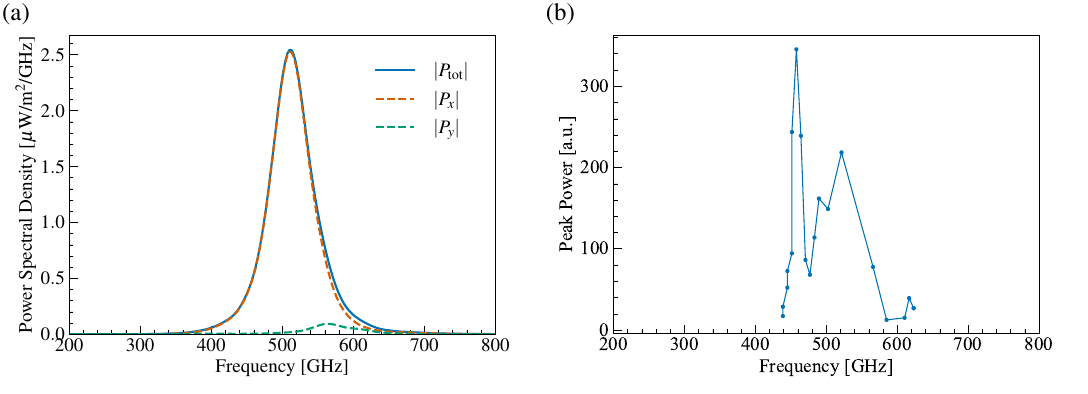}
    \caption[Rectangular mesa (R1): frequency characteristics]{
    (a) Simulated radiated power $P_{\mathrm{sim}}(f)$ integrated over the observation plane (Eq.~\eqref{meq:radiated_power}), and polarization components $\lvert P_x\rvert$ and $\lvert P_y\rvert$ obtained from the Jones vector reconstructed from the Poynting-flux-weighted Stokes parameters (Eqs.~\eqref{meq:s0_at_each_probe}–\eqref{meq:delta}). A single dominant resonance appears in the $\sim\!500$~GHz range.
    (b) Experimental peak-power spectrum versus frequency for Device R1 obtained by sweeping the bias point; unlike the smooth unimodal simulation, the measurement exhibits finer variations in the peak magnitude across frequency.}
    \label{fig:frequency_spectrum_of_device_A}
\end{figure}

The simulation result of the frequency characteristics of device R1 is shown in Figure \ref{fig:frequency_spectrum_of_device_A} (a), and the experimental result is shown in Figure \ref{fig:frequency_spectrum_of_device_A} (b). The experimental results are the frequency characteristics of the radiation peaks obtained by sweeping the bias point. While the simulation yields a smooth distribution, the experiment reveals finer variations in intensity. These finer variations will be discussed briefly in the Discussion.

\begin{figure}[htbp]
    \centering
    \includegraphics[width=\linewidth]{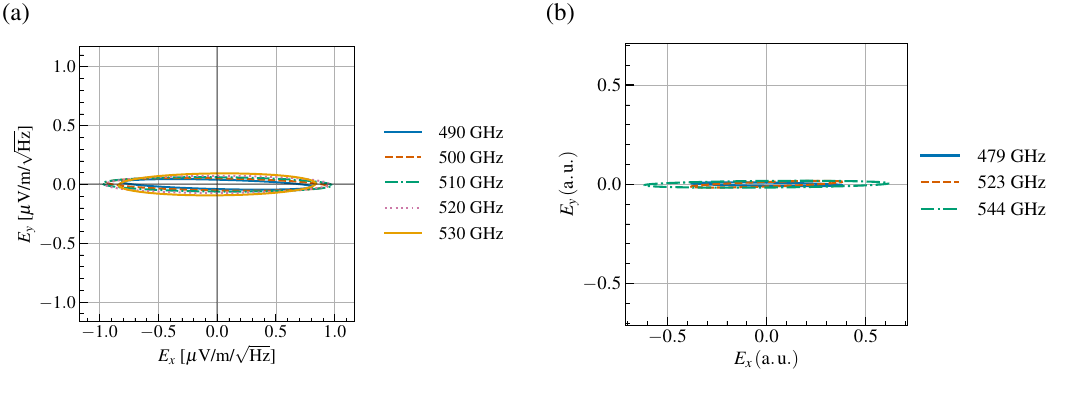}
    \caption[Rectangular mesa (R1): polarization characteristics]{
    (a) Simulated polarization ellipses plotted in the $E_x$–$E_y$ plane at 490, 500, 510, 520, and 530~GHz. The ellipses are highly flattened along the $x$-axis, indicating nearly linear polarization aligned with the mesa's long side (dominant $x$-polarized mode).
    (b) Measured polarization plots at three bias voltages (0.86, 0.93, and 0.96~V) show emission predominantly polarized along $x$, consistent with the simulation.}
    \label{fig:polarization_of_device_A}
\end{figure}

The simulated polarization characteristics are shown in Figure \ref{fig:polarization_of_device_A} (a), and the experimental result is shown in Figure \ref{fig:polarization_of_device_A} (b). In both the simulation and the experiment, polarization characteristics extending in the $x$-direction were observed.

\subsection{Truncated square device}

\begin{figure}[htbp]
    \centering
    \includegraphics[width=\linewidth]{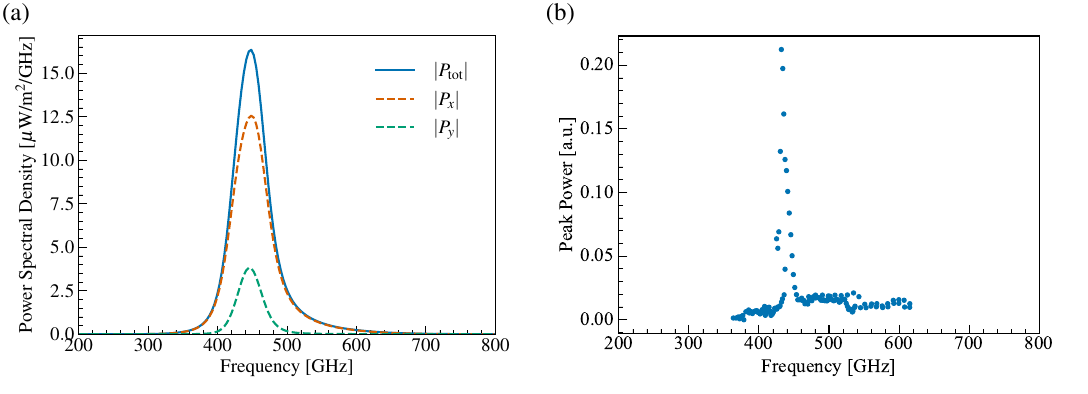}
    \caption[Truncated-square mesa (TS): frequency characteristics]{
    (a) Simulated radiated power for Device TS, showing $P_{\mathrm{sim}}(f)$, $\lvert P_x\rvert$, and $\lvert P_y\rvert$ computed as in Fig.~\ref{fig:frequency_spectrum_of_device_A}(a). A dominant resonance appears in the mid-400~GHz band.
    (b) Experimental peak power versus frequency obtained by sweeping the bias point. Some frequency values are interpolated under a first-order relation between bias current and frequency, following the experimental procedure.}
    \label{fig:frequency_spectrum_of_device_B}
\end{figure}

The simulation result of the frequency characteristics of device TS is shown in Figure \ref{fig:frequency_spectrum_of_device_B} (a), and the experimental result is shown in Figure \ref{fig:frequency_spectrum_of_device_B} (b). The experimental results are the frequency characteristics of the radiation peaks obtained by sweeping the bias point. Some frequencies are interpolated by assuming a first-order relationship between the bias current and the frequency.

\begin{figure}[htbp]
    \centering
    \includegraphics[width=\linewidth]{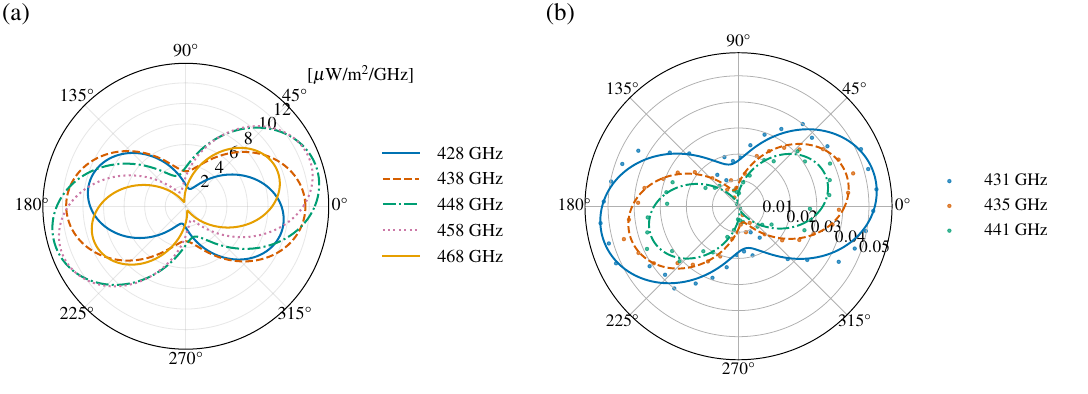}
    \caption[Truncated-square mesa (TS): polarization-angle dependence]{
    (a) Simulated intensity transmitted through an ideal linear polarizer as a function of polarizer angle $\varphi$ for several frequencies (428, 438, 448, 458, and 468~GHz). The near-isotropic angular response around the resonance indicates a low axial ratio (near-circular polarization).
    (b) Measured angular intensity at 431, 435, and 441~GHz likewise exhibits low axial ratio within the emission band, consistent with the simulation trends.}
    \label{fig:polarization_of_device_B}
\end{figure}

The simulated polarization characteristics of device TS are shown in Figure \ref{fig:polarization_of_device_B} (a), and the experimental result is shown in Figure \ref{fig:polarization_of_device_B} (b). Note that Figure \ref{fig:polarization_of_device_B} (a) shows the angular dependence of the polarization \textit{intensity}, not the polarization ellipse of the radiated electric field. The reason for not making a direct comparison of the polarization ellipse is that the Stokes parameters were not measured in the experiment, and the polarization ellipse cannot be reconstructed from the data.

\section{Discussion}
For both devices, the frequency characteristics of the radiation peak intensity exhibit a smoother unimodal profile in simulations, whereas experiments show a more jagged or sharp shape.
This suggests that, in addition to device heating, nonlinear interactions between Josephson junctions destabilize in-phase oscillations even within the resonance range.
In globally coupled massive nonlinear systems, synchronization has been reported to exhibit temporal oscillations \cite{Munyayev2023-yi,Bolotov2024-cl}.
This situation may also occur in JPE, necessitating further analysis with models that incorporate nonlinearity.
Also, device R1 exhibited a flatter polarization characteristic in the experiment compared to the simulation.
This suggests that the horizontally polarized (TM01) mode may have suppressed the circularly polarized mode due to phase locking between the Josephson junctions.
Thus, the discrepancy between simulation and experimental results highlights temperature non-uniformity and electromagnetic nonlinearity in the actual device.

Although this method significantly reproduces the resonance mode excited in the JPE, it is subject to the following constraints because it is a calculation based on approximation. First, because the system's nonlinearity is excluded, it is not possible to reproduce the generation of harmonics, the characteristic electromagnetic field modes, such as the $\pi$-kink, that emerge with it, and the interaction between junctions that leads to frequency locking. In addition, as mentioned above, since the Josephson relation itself is also linearized, it is necessary to give an external excitation to excite the mode. This introduces a slight arbitrariness in selecting the excitation port position in the simulation. For example, in this simulation, we excited the mesa's edge. If we instead excite the center, the mode observed here would not be excited. This is because the modes that couple depend on the location of the excitation port.

On the other hand, this method has the following advantages over conventional LD model-based simulations. First, since it can be run on a commercial electromagnetic field simulator, both flexible modeling and high computational speed can be achieved immediately. In addition, since the nonlinearity of the system is removed, interference between different frequency components does not occur, so it is possible to calculate the wideband response of the JPE at once in the time domain by pulse excitation.

\section{Conclusion}
We developed a three–dimensional, anisotropic dielectric model for intrinsic Josephson-junction stacks by applying a linear approximation and an effective-medium approximation to the Lawrence–Doniach model. In the derivation of the dielectric model, we theoretically demonstrated that the real part of the complex permittivity in the out-of-plane direction varies significantly depending on the bias point.
This suggests potential applications as optical elements in the sub-terahertz range. The resulting complex, dispersive permittivity tensor captures the key linear electrodynamics of Bi$_2$Sr$_2$CaCu$_2$O$_{8+\delta}$, including in-plane inductive (plasma-like) response and the finite-bias reduction of the out-of-plane response to capacitive and normal-loss terms. Implemented as a user-defined material in CST Studio Suite, the model enables full-wave 3D FDTD analysis without custom codes.

Validation against two previously measured mesa devices shows that the approach reproduces the dominant radiation modes and polarization behavior. For a rectangular mesa, the simulated emission frequency and linear polarization agree with experiment. For a truncated-square mesa, the simulated intensity-angle dependence is consistent with the low-axial-ratio emission reported experimentally. Remaining discrepancies in spectral smoothness and polarization contrast are plausibly attributed to thermal inhomogeneity and the junction's nonlinearity, which are intentionally excluded from the present linear model.

Because the method can be implemented in commercial FDTD software, it enables fast parameter sweeps and flexible geometric modeling, making it directly applicable to simulation-driven design of JPEs. In the future, combining this model with a nonlinear electrical circuit model \cite{Kobayashi2022-px,Kobayashi2025-bg}, using the former to extract linear circuit parameters and the latter to evaluate the synchronous-oscillation, will enable more precise prediction of device characteristics.

\section*{SUPPLEMENTARY MATERIAL}
See supplementary material for the derivation of an anisotropic dielectric model based on the axiomatic formulation of the LD model and the numerical analysis of the small-signal response of a Josephson junction in a finite bias state.


\section*{Data Availability Statement}
The data that supports the findings of this study are available from the corresponding author upon reasonable request.

\bibliographystyle{IEEEtran}
\bibliography{terahertz,jpe,references}

@ARTICLE{Kleiner2019-hy,
  title     = "Terahertz emission from
               {Bi}$_{2}${Sr}$_{2}${CaCu}$_{2}${O}$_{8+x}$ intrinsic Josephson
               junction stacks",
  author    = "Kleiner, Reinhold and Wang, Huabing",
  journal   = "J. Appl. Phys.",
  publisher = "AIP Publishing",
  volume    =  126,
  number    =  17,
  pages     =  171101,
  month     =  nov,
  year      =  2019,
  language  = "en"
}

@ARTICLE{Ozyuzer2007-nq,
  title    = "Emission of coherent {THz} radiation from superconductors",
  author   = "Ozyuzer, L and Koshelev, A E and Kurter, C and Gopalsami, N and
              Li, Q and Tachiki, M and Kadowaki, K and Yamamoto, T and Minami, H
              and Yamaguchi, H and Tachiki, T and Gray, K E and Kwok, W-K and
              Welp, U",
  journal  = "Science",
  volume   =  318,
  number   =  5854,
  pages    = "1291--1293",
  month    =  nov,
  year     =  2007,
  language = "en"
}

@ARTICLE{Kashiwagi2015-og,
  title     = "A high-Tc intrinsic Josephson junction emitter tunable from 0.5
               to 2.4 terahertz",
  author    = "Kashiwagi, Takanari and Sakamoto, Kazuki and Kubo, Hiroyuki and
               Shibano, Yuuki and Enomoto, Takuma and Kitamura, Takeo and
               Asanuma, Kentaro and Yasui, Takaki and Watanabe, Chiharu and
               Nakade, Kurama and Saiwai, Yoshihiko and Katsuragawa, Takuya and
               Tsujimoto, Manabu and Yoshizaki, Ryozo and Yamamoto, Takashi and
               Minami, Hidetoshi and Klemm, Richard A and Kadowaki, Kazuo",
  journal   = "Appl. Phys. Lett.",
  publisher = "AIP Publishing",
  volume    =  107,
  number    =  8,
  pages     =  082601,
  month     =  aug,
  year      =  2015
}

@ARTICLE{Kakeya2016-ns,
  title     = "Terahertz-wave emission from {Bi2212} intrinsic Josephson
               junctions: a review on recent progress",
  author    = "Kakeya, Itsuhiro and Wang, Huabing",
  journal   = "Supercond. Sci. Technol.",
  publisher = "IOP Publishing",
  volume    =  29,
  number    =  7,
  pages     =  073001,
  month     =  may,
  year      =  2016,
  language  = "en"
}

@ARTICLE{Miyamoto2024-el,
  title     = "Wide-band frequency modulation of a terahertz intrinsic Josephson
               junction emitter of a cuprate superconductor",
  author    = "Miyamoto, Masashi and Kobayashi, Ryota and Kuwano, Genki and
               Tsujimoto, Manabu and Kakeya, Itsuhiro",
  journal   = "Nat. Photonics",
  publisher = "Nature Publishing Group",
  volume    =  18,
  number    =  3,
  pages     = "267--275",
  month     =  jan,
  year      =  2024,
  language  = "en"
}

@ARTICLE{Elarabi2024-np,
  title     = "Polarized terahertz electromagnetic-wave radiation from cuprate
               superconductor {Bi2212} mesa structures",
  author    = "Elarabi, Asem and Saito, Yoshito and Asai, Hidehiro and
               Kobayashi, Ryota and Hayama, Ken and Maeda, Keiichiro and Fujita,
               Shuma and Yoshioka, Yusuke and Takano, Yoshihiko and Tsujimoto,
               Manabu and Kakeya, Itsuhiro",
  journal   = "Jpn. J. Appl. Phys.",
  publisher = "IOP Publishing",
  volume    =  63,
  number    =  2,
  pages     =  020801,
  month     =  jan,
  year      =  2024,
  language  = "en"
}

@ARTICLE{Kadowaki2008-rs,
  title    = "Direct observation of tetrahertz electromagnetic waves emitted
              from intrinsic Josephson junctions in single crystalline
              {Bi}$_{2}${Sr}$_{2}${CaCu}$_{2}${O}$_{8+\delta}$",
  author   = "Kadowaki, K and Yamaguchi, H and Kawamata, K and Yamamoto, T and
              Minami, H and Kakeya, I and Welp, U and Ozyuzer, L and Koshelev, A
              and Kurter, C and Gray, K E and Kwok, W-K",
  journal  = "Phys. C",
  volume   =  468,
  number   =  7,
  pages    = "634--639",
  month    =  apr,
  year     =  2008
}

@ARTICLE{Tsujimoto2012-le,
  title     = "Terahertz imaging system using high-\textit{Tc} superconducting
               oscillation devices",
  author    = "Tsujimoto, M and Minami, H and Delfanazari, K and Sawamura, M and
               Nakayama, R and Kitamura, T and Yamamoto, T and Kashiwagi, T and
               Hattori, T and Kadowaki, K",
  journal   = "J. Appl. Phys.",
  publisher = "AIP Publishing",
  volume    =  111,
  number    =  12,
  pages     =  123111,
  month     =  jun,
  year      =  2012,
  language  = "en"
}

@ARTICLE{Delfanazari2025-ys,
  title     = "Chip-scale electrically driven superconducting coherent photon
               sources for quantum information processing",
  author    = "Delfanazari, Kaveh",
  journal   = "Nat. Photonics",
  publisher = "Springer Science and Business Media LLC",
  pages     = "1--15",
  month     =  sep,
  year      =  2025,
  language  = "en"
}

@ARTICLE{Elarabi2016-lr,
  title    = "Polarization Enhancement of Terahertz Radiation Generated by
              Intrinsic Josephson Junctions in a Truncated Edge Square
              Bi$_2$Sr$_2$CaCu$_2$O$_{8+\delta}$ Mesa",
  author   = "Elarabi, A and Yoshioka, Y and Tsujimoto, M and Nakagawa, Y and
              Kakeya, I",
  journal  = "Phys. Procedia",
  volume   =  81,
  pages    = "133--136",
  month    =  jan,
  year     =  2016
}

@ARTICLE{Elarabi2017-uk,
  title     = "Monolithic Superconducting Emitter of Tunable Circularly
               Polarized Terahertz Radiation",
  author    = "Elarabi, A and Yoshioka, Y and Tsujimoto, M and Kakeya, I",
  journal   = "Phys. Rev. Appl.",
  publisher = "American Physical Society",
  volume    =  8,
  number    =  6,
  pages     =  064034,
  month     =  dec,
  year      =  2017
}

@ARTICLE{Gross2012-vn,
  title     = "Hot-spot formation in stacks of intrinsic Josephson junctions in
               {Bi}$_{2}${Sr}$_{2}${CaCu}$_{2}${O}$_{8}$",
  author    = "Gross, B and Guénon, S and Yuan, J and Li, M Y and Li, J and
               Ishii, A and Mints, R G and Hatano, T and Wu, P H and Koelle, D
               and Wang, H B and Kleiner, R",
  journal   = "Phys. Rev. B Condens. Matter",
  publisher = "American Physical Society",
  volume    =  86,
  number    =  9,
  pages     =  094524,
  month     =  sep,
  year      =  2012
}

@ARTICLE{Gross2015-zi,
  title     = "Electrothermal behavior and terahertz emission properties of a
               planar array of two {Bi}$_{2}${Sr}$_{2}${CaCu}$_{2}${O}$_{8+\delta}$
               intrinsic Josephson junction stacks",
  author    = "Gross, B and Rudau, F and Kinev, N and Tsujimoto, M and Yuan, J
               and Huang, Y and Ji, M and Zhou, X J and An, D Y and Ishii, A and
               Wu, P H and Hatano, T and Koelle, D and Wang, H B and Koshelets,
               V P and Kleiner, R",
  journal   = "Supercond. Sci. Technol.",
  publisher = "IOP Publishing",
  volume    =  28,
  number    =  5,
  pages     =  055004,
  month     =  mar,
  year      =  2015,
  language  = "en"
}

@ARTICLE{Hu2009-yo,
  title     = "Cavity phenomena in mesas of cuprate high-{T}$_{c}$
               superconductors under voltage bias",
  author    = "Hu, Xiao and Lin, Shizeng",
  journal   = "Phys. Rev. B Condens. Matter",
  publisher = "American Physical Society",
  volume    =  80,
  number    =  6,
  pages     =  064516,
  month     =  aug,
  year      =  2009
}

@ARTICLE{Koyama2009-an,
  title     = "In-phase electrodynamics and terahertz wave emission in extended
               intrinsic Josephson junctions",
  author    = "Koyama, Tomio and Matsumoto, Hideki and Machida, Masahiko and
               Kadowaki, Kazuo",
  journal   = "Phys. Rev. B Condens. Matter",
  publisher = "American Physical Society",
  volume    =  79,
  number    =  10,
  pages     =  104522,
  month     =  mar,
  year      =  2009
}

@ARTICLE{Welp2013-eb,
  title     = "Superconducting emitters of {THz} radiation",
  author    = "Welp, Ulrich and Kadowaki, Kazuo and Kleiner, Reinhold",
  journal   = "Nat. Photonics",
  publisher = "Nature Publishing Group",
  volume    =  7,
  number    =  9,
  pages     = "702--710",
  month     =  aug,
  year      =  2013,
  language  = "en"
}

@ARTICLE{Yurgens2011-fr,
  title     = "Temperature distribution in a large
               {Bi}$_{2}${Sr}$_{2}${CaCu}$_{2}${O}$_{\delta}$ mesa",
  author    = "Yurgens, A",
  journal   = "Phys. Rev. B Condens. Matter",
  publisher = "American Physical Society",
  volume    =  83,
  number    =  18,
  pages     =  184501,
  month     =  may,
  year      =  2011
}

@ARTICLE{Rakhmanov2010-xp,
  title     = "Layered superconductors as negative-refractive-index
               metamaterials",
  author    = "Rakhmanov, A L and Yampol'skii, V A and Fan, J A and Capasso,
               Federico and Nori, Franco",
  journal   = "Phys. Rev. B Condens. Matter",
  publisher = "American Physical Society",
  volume    =  81,
  number    =  7,
  pages     =  075101,
  month     =  feb,
  year      =  2010
}

@ARTICLE{Poddubny2013-wg,
  title     = "Hyperbolic metamaterials",
  author    = "Poddubny, Alexander and Iorsh, Ivan and Belov, Pavel and Kivshar,
               Yuri",
  journal   = "Nat. Photonics",
  publisher = "Nature Publishing Group",
  volume    =  7,
  number    =  12,
  pages     = "948--957",
  month     =  nov,
  year      =  2013,
  language  = "en"
}

@ARTICLE{Smith2003-ji,
  title    = "Electromagnetic wave propagation in media with indefinite
              permittivity and permeability tensors",
  author   = "Smith, D R and Schurig, D",
  journal  = "Phys. Rev. Lett.",
  volume   =  90,
  number   =  7,
  pages    =  077405,
  month    =  feb,
  year     =  2003,
  language = "en"
}

@ARTICLE{Rudau2016-ov,
  title     = "Three-Dimensional Simulations of the Electrothermal and Terahertz
               Emission Properties of {Bi}$_{2}${Sr}$_{2}${CaCu}$_{2}${O}$_{8}$
               Intrinsic Josephson Junction Stacks",
  author    = "Rudau, F and Wieland, R and Langer, J and Zhou, X J and Ji, M and
               Kinev, N and Hao, L Y and Huang, Y and Li, J and Wu, P H and
               Hatano, T and Koshelets, V P and Wang, H B and Koelle, D and
               Kleiner, R",
  journal   = "Phys. Rev. Appl.",
  publisher = "American Physical Society",
  volume    =  5,
  number    =  4,
  pages     =  044017,
  month     =  apr,
  year      =  2016
}

@ARTICLE{Hu2010-tv,
  title     = "Phase dynamics in a stack of inductively coupled intrinsic
               Josephson junctions and terahertz electromagnetic radiation",
  author    = "Hu, Xiao and Lin, Shi-Zeng",
  journal   = "Supercond. Sci. Technol.",
  publisher = "IOP Publishing",
  volume    =  23,
  number    =  5,
  pages     =  053001,
  month     =  apr,
  year      =  2010,
  language  = "en"
}

@ARTICLE{Machida2010-jw,
  title     = "Theory and simulation of electromagnetic wave emission from
               intrinsic Josephson junctions",
  author    = "Machida, M and Ota, Y and Sasa, N and Koyama, T and Matsumoto, H",
  journal   = "J. Phys. Conf. Ser.",
  publisher = "IOP Publishing",
  volume    =  248,
  pages     =  012037,
  month     =  nov,
  year      =  2010
}

@ARTICLE{Kobayashi2022-px,
  title     = "Spontaneous Frequency Shift and Phase Delay of Coupled Terahertz
               Radiation Mediated by the Josephson Plasmon in a Cuprate
               Superconductor",
  author    = "Kobayashi, Ryota and Hayama, Ken and Fujita, Shuma and Tsujimoto,
               Manabu and Kakeya, Itsuhiro",
  journal   = "Phys. Rev. Appl.",
  publisher = "American Physical Society",
  volume    =  17,
  number    =  5,
  pages     =  054043,
  month     =  may,
  year      =  2022
}

@ARTICLE{Saito2022-fw,
  title     = "Polarization analysis of terahertz emission from Bi-2212
               cross-whisker intrinsic Josephson junction devices and their
               refractive index",
  author    = "Saito, Y and Kakeya, I and Takano, Y",
  journal   = "Appl. Phys. Lett.",
  publisher = "AIP Publishing",
  volume    =  121,
  number    =  21,
  pages     =  212601,
  month     =  nov,
  year      =  2022,
  language  = "en"
}

@ARTICLE{Kobayashi2025-bg,
  title     = "Extensible approach to nonlinear circuit simulation of terahertz
               radiation phenomena from stacked Josephson junctions",
  author    = "Kobayashi, R and Kakeya, I",
  journal   = "Supercond. Sci. Technol.",
  publisher = "IOP Publishing",
  volume    =  38,
  number    =  2,
  pages     = "02LT01",
  month     =  feb,
  year      =  2025
}

@ARTICLE{Lawrence1971-tv,
  title     = "Theory of layer-structure superconductors",
  author    = "Lawrence, W E and Doniach, S",
  journal   = "Proceedings of the Twelfth International Conference on Low
               Temperature Physics",
  publisher = "Tokyo Keigaku Publishing Co., Ltd.",
  pages     = "361--362",
  year      =  1971
}

@ARTICLE{Bolotov2024-cl,
  title     = "Breathing and switching cyclops states in Kuramoto networks with
               higher-mode coupling",
  author    = "Bolotov, Maxim I and Munyayev, Vyacheslav O and Smirnov, Lev A
               and Osipov, Grigory V and Belykh, Igor",
  journal   = "Phys. Rev. E.",
  publisher = "American Physical Society (APS)",
  volume    =  109,
  number    = "5-1",
  pages     =  054202,
  month     =  may,
  year      =  2024,
  language  = "en"
}

@ARTICLE{Munyayev2023-yi,
  title     = "Cyclops states in repulsive Kuramoto networks: The role of
               higher-order coupling",
  author    = "Munyayev, Vyacheslav O and Bolotov, Maxim I and Smirnov, Lev A
               and Osipov, Grigory V and Belykh, Igor",
  journal   = "Phys. Rev. Lett.",
  publisher = "American Physical Society (APS)",
  volume    =  130,
  number    =  10,
  pages     =  107201,
  month     =  mar,
  year      =  2023,
  language  = "en"
}

\newpage

 



\begin{IEEEbiographynophoto}{Ryota Kobayashi}
received the B.Eng. degree in electrical and electronic engineering in 2021 and the M.Eng. and Ph.D. degrees in electronic science and engineering in 2023 and 2026, respectively, from Kyoto University, Kyoto, Japan.\par He was a JSPS Research Fellow (DC2) from 2024 to 2026. His doctoral research focused on mathematical modeling of high-temperature cuprate superconducting terahertz sources for enhanced radiation power. In April 2026, he joined PKSHA Technology Inc. as an Algorithm Engineer, working on real-world applications of machine learning algorithms.\par Dr. Kobayashi received the JSAP Young Scientist Presentation Award for the 2024 Autumn Meeting and the Best Student Oral Presentation Award at APROPOS19 in 2024.
\end{IEEEbiographynophoto}

\begin{IEEEbiographynophoto}{Itsuhiro Kakeya}
received the B.Sci. degree in 1994 and the M.Sci. and Ph.D. degrees in physics in 1996 and 1998, respectively, from Osaka University, Japan.\par He was a research associate and lecturer in University of Tsukuba between 1998 and 2008. He moved to Kyoto University at 2008 as an associate Professor.\par His research subject is electronics-oriented low temperature physics in solids. 
\end{IEEEbiographynophoto}

\end{document}